\documentclass{article}

\usepackage{arxiv}

\usepackage[utf8]{inputenc} 
\usepackage[T1]{fontenc}    
\usepackage{hyperref}       
\usepackage{url}            
\usepackage{booktabs}       
\usepackage{amsfonts}       
\usepackage{nicefrac}       
\usepackage{microtype}      
\usepackage{lipsum}

\usepackage[english]{babel}
\usepackage[square, numbers, sort&compress]{natbib}
\newtheorem{theorem}{Theorem}[section]

\newtheorem{example}[theorem]{Example}

\usepackage{amssymb}
\usepackage{eqnarray,amsmath}
   
 \usepackage{amsmath}
\DeclareMathOperator{\sech}{sech}

\usepackage{float}
\usepackage{subfigure}
\usepackage{graphicx}

\title{Homotopy Analysis Technique for a Generalised (1+1)-Dimensional KdV Equation of Variable Coefficients}
\author{
  Ali Joohy\\
  Department of Computer Science and Mathematics\\
  Kufa University\\
  \texttt{alijoohy@gmail.com}\\
}

\begin{document}
\maketitle

\begin{abstract}
In this work, an exact solution to a new generalized nonlinear KdV partial differential equations has been investigated using homotopy analysis techniques. The mentioned partial differential equation has been solved using homotopy perturbation method (HPM). In details, the study is divided into two cases; the first case is that the linear part is the velocity (the first derivative with respect to time), and the initial guess was chosen at the initial time or the boundary values. Next, the other case was that the linear part consists of many linear terms from the equation, the chosen terms constructed a  partial differential equation whose solution can be gained easily. Eventually, some plots of the found solutions have demonstrated the convergent behaviors of the solutions.
\end{abstract}


\section{Introduction}
\label{sec1}
The finding of the analytical solutions of the nonlinear partial differential equations is one of the most important subjects in modern science; it plays a significant role in various applications in physics and engineering, such equations and their solutions describe problems in biology, aerodynamic, oceanography, meteorology, electronics, and cosmology as well. Korteweg-de Vries (KdV) equations have taken a noticeable amount of scientists’ attention since it had been discovered, in particular, in modeling the free-surface flow in shallow water \cite{katopodes2018free}, and in fluid mechanics applications \cite{lustri2012free,binder2005forced,binder2005free,binder2013free,binder2006steady,wade2017steep}, in general.

The generalized KdV equations are a general form for many kinds of nonlinear partial differential equations that satisfy the famous class of Korteweg-de Vries equations. Many methods and techniques had been discovered, and others have been developed to find the solutions of nonlinear partial differential equations. Computational mathematics and symbolic programming have heightened the evolution of solving nonlinear problems, and it has become easier to handle and rearrange the sophisticated mathematical equations. Most of the complicated nonlinear partial differential equations were solved by numerical analysis \cite{zhang2018numerical}, and asymptotic perturbation techniques \cite{lan2018asymptotic}. The asymptotic perturbation technique has some limitations, that is the parameter $\epsilon$ must be very small so that perturbed term can be neglected as opposed to HPM method which has more flexibility. Recently, many researchers have solved the classic form of  KdV equations by numerical methods and approximated analytical methods \cite{zhang2018numerical,yokus2018numerical,goswami2017numerical,lakestani2017numerical,samokhin2017nonlinear,sarboland2015numerical}.

\section{Basic Idea of the Homotopy Analysis Techniques}
\label{sec2}
The homotopy analysis techniques are based on the concept of homotopy mapping in topology. The homotopy mapping deforms an object to another object continuously. 
S.Liao considered the differential equations as two objects \cite{liao2012homotopy,liao2013advances}. The linear part (the first object) represents the domain of the homotopy mapping, and the nonlinear part (the second object) represents the range of the same mapping.  When the parameter $0 \leq p \leq 1$ travels from zero to unity the homotopy transfers every point in object one, say $O_{1}$, to object two, say $O_{2}$.

Let's construct the homotopic equation as follows:
$$H= (1-p)\, O_{1} + p\,O_{2},$$
note that, when $p=0$ the homotopic equation gives $H=O_{1}$ and when $p=1$ gives $H=O_{2}$, so when $p$ travels from zero to unity continuously the homotopic transforms $O_{1}$ to $O_{2}$ continuously.

\section{Homotopy Perturbation Method (HPM)}
\label{sec3}
The homotopy perturbation method (HPM) was first published by Dr. He in 1999 \cite{he1999homotopy},and well explained and given some guidelines by Esmail Babolian, A Azizi, and J Saeidian in 2009 \cite{babolian2009some}, to illustrate the basic concept of the method, consider the general nonlinear differential equation
\begin{equation}
\label{eq:main}
A(y(r, t))- f(r, t)=0, 
\end{equation}

with the boundary conditions

\begin{equation}
\label{BC}
B\left( u, \frac{\partial{u} }{\partial{n}} \right), \hspace{0.5cm} r \in \partial \Omega
\end{equation}

where $A$ is the general differential operator, $y(r, t)$ is an unknown function, $r$ and $t$ denote the spatial and the temporal independent variables, $B$ is the boundary operator, $f(r, t)$ is an analytic function, and $\partial \Omega$ is the boundary of the domain $\Omega$.
Generally speaking, the operator $A$ can be divided into two operators, $L$ and $N$, $L$ is the linear operator and $N$ is the nonlinear operator. So the operator $A$ can be written as $A=L+N$ and equation (\ref{eq:main}) can be written as

\begin{equation}
\label{eq:LN}
L(u)+ N(u)- f(r)=0. 
\end{equation}

Based on the homotopy concept one can construct the homotopy mapping 
$$v(r, p): \Omega \times [0,1] \to \mathbb{R}$$
that satisfies

\begin{equation}
\label{hom-1}
H(v(r, t; p), p)= (1-p) [L(v) - L(v_{0})]+ p [A(v)-f(r, t)]=0, \hspace{0.5cm} p \in [0,1]
\end{equation}
where $p$ is the embedded parameter and $v_{0}(r, t)$ is the initial guess for equation (\ref{eq:main}) which ,in term, satisfies the boundary conditions in equation (\ref{BC}). Equation (\ref{hom-1}) is called \emph{the homotopy equation}. In fact, it can be written as follows:

\begin{equation}
\label{hom-2}
H(r, p)=L(v(r, t; p))-L(v_{0}(r, t))+ p \left\{ N(v(r, t; p))+ L(v_{0}(r, t)) \right\}=0.
\end{equation}
It is obvious that:

\begin{align}
\label{inisol}
 p=0 & \Rightarrow   H(v(r, t; 0), 0)=L(v(r, t; 0)) - L(v_{0}(r, t))=0,  \\
 \label{finsol}
  p=1 & \Rightarrow    A(v(r, t; 1))- f(r, t)=0.
\end{align}
Actually, the equation (\ref{finsol}) is the nonlinear differential equation (\ref{eq:main}) itself with solution $y(r, t)$, while, equation (\ref{inisol}) has the solution $v_{0}(r, t)$ as one of its solutions. Note, if the operator $L$ is assumed to be a linear operator the solution $v_{0}(r, t)$ is the only solution we have. So we get the following results
\begin{align}
 v(r, t; 0)&=v_{0}(r, t),  \\
v(r, t; 1)&=y(r, t).
\end{align}
The procedure of the traveling of the embedded parameter is resulted from the deformation of the linear
\footnote{We call it linear as the first name (Dr. He) did but in fact, it could be nonlinear but easy to handle the first guessed solution.}
part to the nonlinear part, from another perspective, the process deforms the solution $v_{0}(r, t)$ to the desired solution $y(r, t)$.

 If the embedded parameter $p$ is \emph{''small parameter''}, based on the concept of the classic perturbation method 
\cite{bender2013advanced}, so the solution can be represented as the infinite series
 \begin{equation}
\label{infser-1}
v(r, t; p) = u_{0}(r, t)+u_{1}(r, t) p +u_{2}(r, t)p^{2}+...,
\end{equation}
by substituting equation (\ref{infser-1}) in equation (\ref{eq:main}) and equating the terms of the same power of $p$, a system of linear equations will be gained, then, by solving for the terms $u_{i}$, equation (\ref{infser-1}) can be determined, after that, by making  the parameter $p$ goes to unity, one has
\begin{equation}
\label{infser-2}
y(r, t) = u_{0}(r, t)+u_{1}(r, t)  +u_{2}(r, t)+...,
\end{equation}
which is the approximated solution to equation (\ref{eq:main}), one can see that this series is convergent in most cases and can be summed to obtain the exact solution, once the summation of the solution is not possible for some reason, one can truncate the series for an appropriate accuracy.

\section{Analyzing the Solution of a Generalized KdV Equation by HPM}
\label{sec4}
Consider the nonlinear KdV partial differential equation
\begin{equation}
\label{mpde}
u_{t} + a(t)\,u\, u_{x} + b(t) \, u+ c(t) \, u_{xxx} = 0,
\end{equation}
As an interpretation of the equation (\ref{mpde}), the first term represents the evolution and the third term represent the linear damping while the fourth term refers to the dispersion and the second term is the \emph{nonlinear} term. Moreover, $a(t)$, $b(t)$  and $c(t)$ are analytic functions on some domain represents the time-dependence coefficients of the nonlinear term, the damping terms, and the dispersive term respectively.

To solve equation (\ref{mpde}) by HPM, the linear part and the initial guess must be determined so that the next terms can be found according to them.

\section{Case One}
In this case, the linear part will be the term $u_{t}$ and the other terms will act as the nonlinear parts, consider the initial guess as $v_{0}= \alpha(x, t)$ for now.

Let $L(\phi)=\phi_{t}$ and $N(\phi)=a(t) \, \phi \, \phi_{x} + b(t) \, \phi+ c(t) \, \phi_{xxx}$ then we construct the homotopy as follow

\begin{equation}
\label{homex1}
\phi_{t} - {v_{0}}_{t} + p \left\{ {v_{0}}_{t}+a(t)\, \phi \, \phi_{x} + b(t) \, \phi+ c(t) \, \phi_{xxx}\right\}=0.
\end{equation}

Assume that $\phi = \sum_{i=0}^{\infty} v_{i} p^{i}$, by substituting it in equation (\ref{homex1}) and using equation (\ref{infser-1})  then equating the terms with the same power of $p$, we get the following system of equations,

\begin{align*}
p^{0}:v_{0}&=\alpha(x, t)\\
p^{1}:v_{1}&=- \int{ \left\{ a(t)\,(v_{0} {v_{0}}_{x})+b(t) v_{0}+c(t){v_{0}}_{xxx} \right\} }\,dt,\\
p^{2}:v_{2}&=- \int{ \left\{ a(t)\,(v_{0} {v_{1}}_{x}+v_{1} {v_{0}}_{x})+b(t) v_{1}+c(t){v_{1}}_{xxx} \right\}}\,dt,
\end{align*}
 by finding more terms the pattern is clearly
 
 \begin{equation}
 \label{finalterms}
v_{i}= -\int \left\{a(t)\,  (\sum_{j=0}^{n=i-1} v_{j} {v_{n-j}}_{x}  ) +b(t) \, v_{i-1} + c(t)\, {v_{i-1}}_{xxx} \right\}\,dt.
\end{equation}
In the general $i'th$ term formula (\ref{finalterms}) all the terms of the series have been determined, so the solution of the generalized equation (\ref{mpde}) has found.

\subsection{Some Numeric Examples}
\begin{figure}[H]
\centering
  \includegraphics[width=10cm]{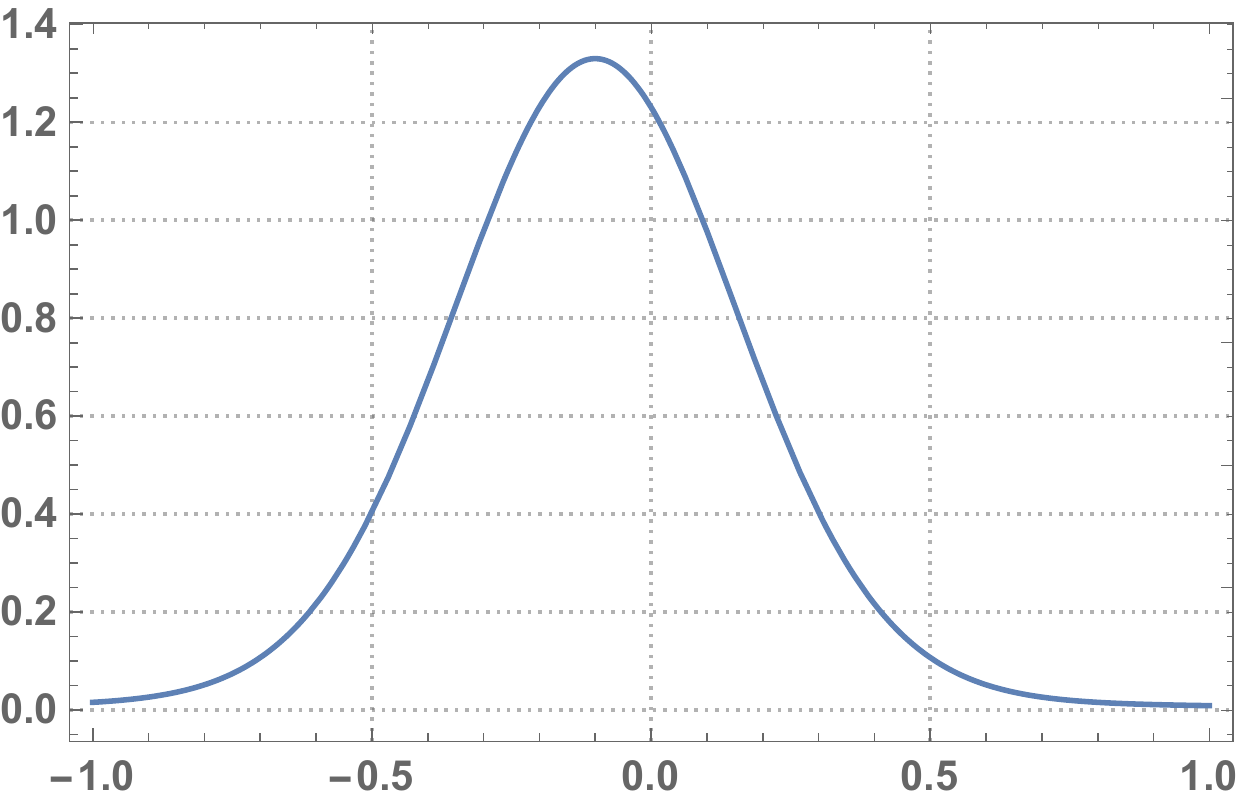}
  \caption{The time shot (t=1) plot of the truncated solution of the equation in example (\ref{ex1}) with the following values of the parameters $r=0.03$, $k=1$ and $m=0.1$.}
  \label{fig:1}
\end{figure}

\begin{example}
\label{ex1}
Lets take the following coefficients $a(t)=6$, $b(t)=0$ and $c(t)=constant$, and the boundary conditions $u_{xxx} \to 0$ as $t \to \infty$, and the initial guess will be the profile $v_{0}(x, t)=r \, \sech^{2}{k\, (x-t\,m)} $ where the movement of the wave is steady. After tedious calculations the terms of the series are

\begin{align*}
v_{0}(x, t)&=r \, \sech^{2}{k\, (x-t\,m)},\\
v_{1}(x, t)&=\frac{3\, r^{2} \, \sech^{4}{k\, (x-t\,m)}}{m},\\
v_{2}(x, t)&=\frac{13\, r^{3} \, \sech^{6}{k\, (x-t\,m)}}{m^{2}},\\
v_{3}(x, t)&=\frac{171\, r^{4} \, \sech^{8}{k\, (x-t\,m)}}{2m^{3}} + \frac{507\, r^{6} \, \sech^{12}{k\, (x-t\,m)}}{m^{5}},\\
v_{4}(x, t)&=\frac{747\, r^{5} \, \sech^{10}{k\, (x-t\,m)}}{m^{4}} + \frac{3042\, r^{7} \, \sech^{14}{k\, (x-t\,m)}}{m^{6}},\\
v_{5}(x, t)&=\frac{6528\, r^{4} \, \sech^{12}{k\, (x-t\,m)}}{m^{5}} + \frac{27378\, r^{8} \, \sech^{12}{k\, (x-t\,m)}}{m^{7}},\\
v_{6}(x, t)&=\frac{59283\, r^{4} \, \sech^{14}{k\, (x-t\,m)}}{m^{6}} + \frac{258570\, r^{9} \, \sech^{18}{k\, (x-t\,m)}}{m^{8}},
\end{align*}

five terms of the series have been found using the general $i'th$ formula (\ref{finalterms}) and the initial and boundary conditions. The final approximated solution after the truncation and the calculating of first four terms must be 

\begin{equation}
u(x, t)= v_{0}(x, t)+v_{1}(x, t)+v_{2}(x, t)+v_{3}(x, t)+v_{4}(x, t)+v_{5}(x, t)+...,
\end{equation}
adding the six terms and simplify the summation gives
\begin{align*}
u(x, t)&= \frac{258570 r^9 \, \sech^{18}(k (x-m t))}{m^8}+\frac{27378 r^8 \, \sech^{16}(k (x-m t))}{m^7}\\
&+\frac{62325 r^7 \, \sech^{14}(k (x-m t))}{m^6}+\frac{7035 r^6 \, \sech^{12}(k (x-m t))}{m^5}\\
&+\frac{747 r^5 \, \sech^{10}(k (x-m t))}{m^4}+\frac{171 r^4 \, \sech^8(k (x-m t))}{2 m^3},\\
&+\frac{13 r^3 \, \sech^6(k (x-m t))}{m^2}+\frac{3 r^2 \, \sech^4(k (x-m t))}{m}+r \, \sech^2(k (x-m t)),
\end{align*}
where $r$ is the amplitude of the wave, $k$ refers to the number of waves and $m$ is the speed of the wave, One can change the values of the parameters of the initial guess to gain the suitable form of solution to the problem he studies, see figure (\ref{fig:1}).
\end{example}

\begin{figure}[H]
\centering
  \includegraphics[width=10cm]{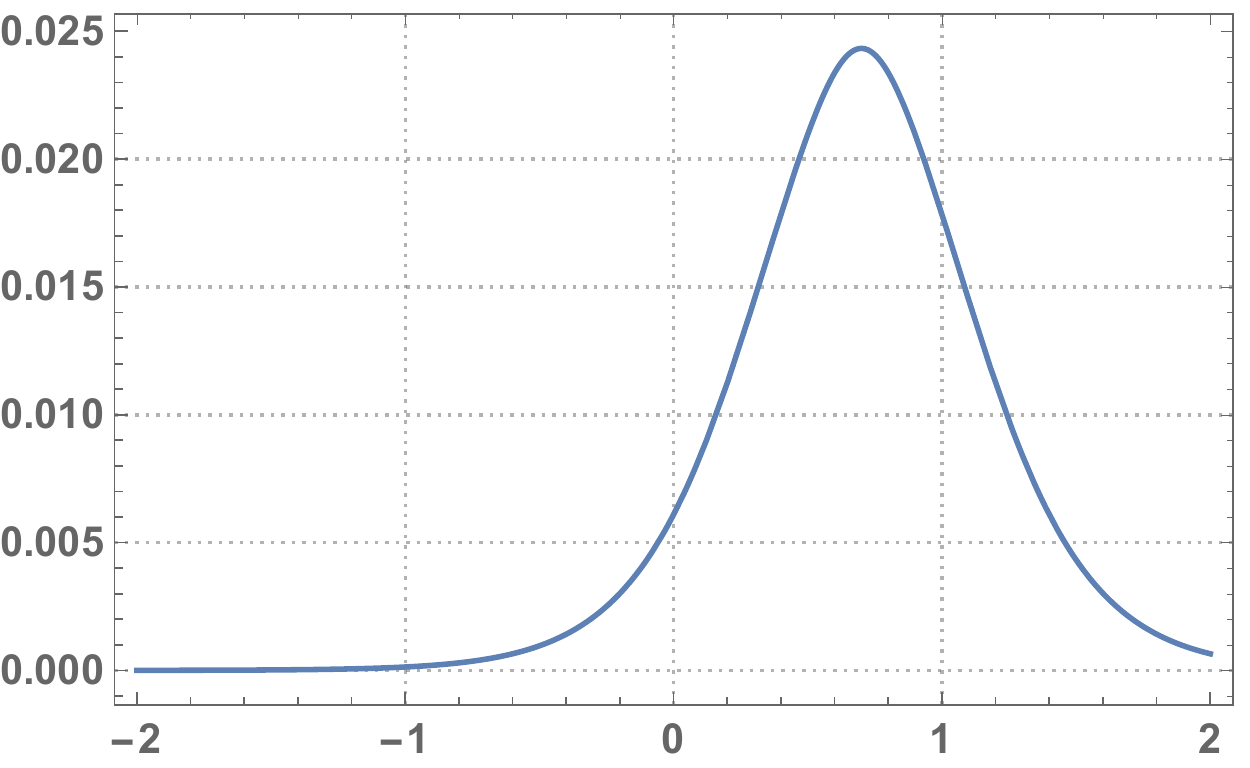}
  \caption{The time shot (t=1) plot of the truncated solution of the equation in example (\ref{ex2}) with the following values of the parameters $A=0.03$, $B=2$, $C=0.6$ and $D=-2$.}
  \label{fig:2}
\end{figure}

\begin{example}
\label{ex2}
Let's take the same problem in (\ref{ex1}) but with the initial guess 
$$u(x, t)=A\, \sech^2(B x+C t+D)$$ 
where $A$, $B$, $C$ and $D$ are physical constants, by appling the same algorithm, the terms of the series solution are
\begin{align*}
v_{0}(x, t)&=A\, \text{sech}^2(B x+C t+D)\\
v_{1}(x, t)&=-\frac{3 A^2 B\, \text{sech}^4(B x+C t+D)}{C},\\
v_{2}(x, t)&=\frac{13 A^3 B^2 \, \text{sech}^6(B x+C t+D)}{C^2},\\
v_{3}(x, t)&=-\frac{507 A^6 B^5 \text{sech}^{12}(B x+C t+D)}{C^5}-\frac{171 A^4 B^3 \, \text{sech}^8(B x+C t+D)}{2 C^3},\\
v_{4}(x, t)&=\frac{3042 A^7 B^6 \, \text{sech}^{14}(B x+C t+D)}{C^6}+\frac{747 A^5 B^4\, \text{sech}^{10}(B x+C t+D)}{C^4},\\
v_{5}(x, t)&=-\frac{27378 A^8 B^7\,\text{sech}^{16}(B x+C t+D)}{C^7}-\frac{6528 A^6 B^5\,\text{sech}^{12}(B x+C t+D)}{C^5},\\
v_{6}(x, t)&=\frac{258570 A^9 B^8\, \text{sech}^{18}(B x+C t+D)}{C^8}+\frac{59283 A^7 B^6\, \text{sech}^{14}(B x+C t+D)}{C^6}
\end{align*}
six terms will give a tremendous accuracy, one can go on further to find more terms, see figure (\ref{fig:2}).
\end{example}

\section{Case Two}
\label{casetwo}
As it has been mentioned in section (\ref{sec1}), the linear part does not have to be precisely \emph{linear}, it could be nonlinear but easy to handle its solution. 

Let's take the first and the third terms in equation (\ref{mpde}) as the linear part, so that
 $$L(\phi)= \phi_{t} + b(t)\, \phi,$$ 
 and the nonlinear part 
 $$N(\phi) = a(t)\, \phi \, \phi_{x} + c(t)\, \phi_{xxx}.$$
After solving the linear part, it is easy to see that the initial solution profile will be the solution to the linear part, 
\begin{equation}
\label{initialsol}
v_{0}(x, t)=f(x)\, e^{-\int {b(t)\, dt}}.
\end{equation}
where $f(x)$ is a function for the spatial variable.

Using equation (\ref{hom-2}) the homotopic will be as follows
\begin{equation}
\label{homc2}
\phi_{t} + b(t)\, \phi- ({v_{0}}_{t} + b(t)\, v_{0} )+ p \{ {v_{0}}_{t} + b(t)\, v_{0} + a(t)\, \phi \, \phi_{x} + c(t)\, \phi_{xxx}\}=0,
\end{equation}

by assuming the solution $\phi$ can be represented by a power series for the parameter $p$, $\phi$ can be written as

\begin{equation}
\label{ser2}
\phi (v, p) = v_{0} + v_{1} \, p+ v_{2}\, p^{2} + ...,
\end{equation}
substituting the series (\ref{ser2}) into the equation (\ref{homc2}) and equate the terms with the same power of $p$ the values of $v_{i}$ will be determined, then, the series terms will be found. Eventually, the final solution $u(x, t)$ is 

\begin{equation}
u(x, t)= \lim_{p \to 1}{\phi(x, ,t)}.
\end{equation}
After equating the terms with the same power of $p$, the result is a system of forced ODEs for the variables $u, t$.

\begin{align*}
p^{0}:&v_{0}=\alpha(x, t)\\
p^{1}:&{v_{1}}_{t} + b(t)\, v_{1} = - \{ {v_{0}}_{t} + b(t)\, v_{0} + a(t)\, [v_{0}\, {v_{0}}_{x}] + c(t)\, {v_{0}}_{xxx} \},\\
p^{2}:&{v_{2}}_{t} + b(t)\, v_{2} = - \{  a(t)\, [v_{0}\, {v_{1}}_{x}+v_{1}\, {v_{0}}_{x}] + c(t)\, {v_{1}}_{xxx} \},\\
p^{3}:&{v_{3}}_{t} + b(t)\, v_{3} = - \{ a(t)\, [v_{0}\, {v_{2}}_{x}+v_{1}\, {v_{1}}_{x}+v_{2}\, {v_{0}}_{x}] + c(t)\, {v_{2}}_{xxx} \},\\
\end{align*}
the pattern is easily seen,
 
 \begin{equation}
 \label{proposedalg}
 {v_{i}}_{t} + b(t)\, v_{i}= - \{ a(t) \, ( \sum_{j=0}^{n=i-1} v_{j} {v_{n-j}}_{x} ) \}, \hspace{0.5cm} i\geq2.
\end{equation}
In the presiding work, the unforced nonlinear problem has been simplified to a non-finite set of linear forced problems, so by solving these problems, the terms of the series will construct an approximate solution to the problem (\ref{homex1}).

\subsection{Some Numeric Examples}
\begin{figure}[H]
\centering
  \includegraphics[width=10cm]{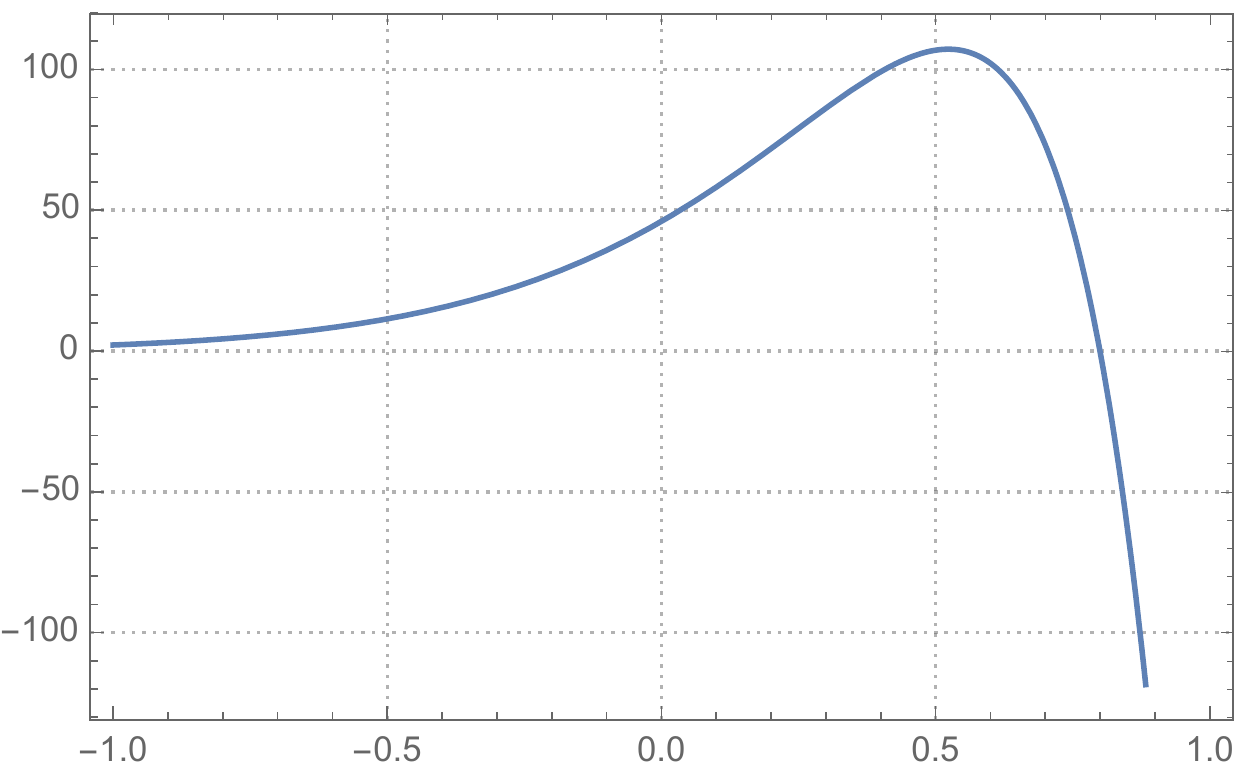}
  \caption{The time shot (t=1) plot of the truncated solution of the equation in example (\ref{ex2-1}) in section (\ref{casetwo}) with the following values of the coefficients  $a(t)=3$, $b(t)=2$ and $c(t)=-1$.}
  \label{fig:3}
\end{figure}

\begin{example}
\label{ex2-1}
In this example, an investigation for the following equation will be revealed
\begin{equation}
u_{t} + 3\,u u_{x} + 2 \, u - \, u_{xxx} = 0,
\end{equation}
first, the linear part is $u_{t}+ 2\, u=0$ the initial solution must be calculated by (\ref{initialsol}), 
\begin{equation}
\label{Gini}
v_{0}(x, t)= f(x)\,e^{- \int 2\, dt},
\end{equation}
$f(x)$ is a function of the variable $x$,here the author chose $e^{x}$, so assume that  $v_{0}(x, t)= e^{x-2t}$ and at $t=0$  the initial solution is $v_{0}(x, 0)=  e^{x}$, by applying the proposed algorithm in $i'th$ term formula (\ref{proposedalg}) the first three terms were found as:
\begin{align}
v_{0}(x, t) &= e^{x}, \\
v_{1}(x, t) &= e^{-2 t}-\frac{1}{2} e^x \left(3 e^x+1\right),\\
v_{2}(x, t) &= \frac{1}{2} e^{x-2 t} \left(-9 e^{2 t+x}+\frac{27}{2} e^{2 t+2 x}-6 t-\frac{e^{2 t}}{2}\right)+e^{-2 t},\\
v_{3}(x, t)&= e^{-2 t}-\frac{1}{4} e^{x-2 t} \\
                &\times \left(-36 t^2 e^x+6 t^2-36 t e^x+\frac{141}{2} e^{2 t+x}-432 e^{2 t+2 x} +189 e^{2 t+3 x}+6 t+\frac{e^{2 t}}{2} \right),
\end{align}
then
\begin{equation}
   u(x, t) = v_{0}(x, t)+v_{1}(x, t)+v_{2}(x, t)+v_{3}(x, t)+...,
\end{equation}

It is possible that one finds more terms for better accuracy, see figure(\ref{fig:3}).
\end{example}

\begin{figure}[H]
\centering
  \begin{subfigure}
    \centering\includegraphics[width=6cm]{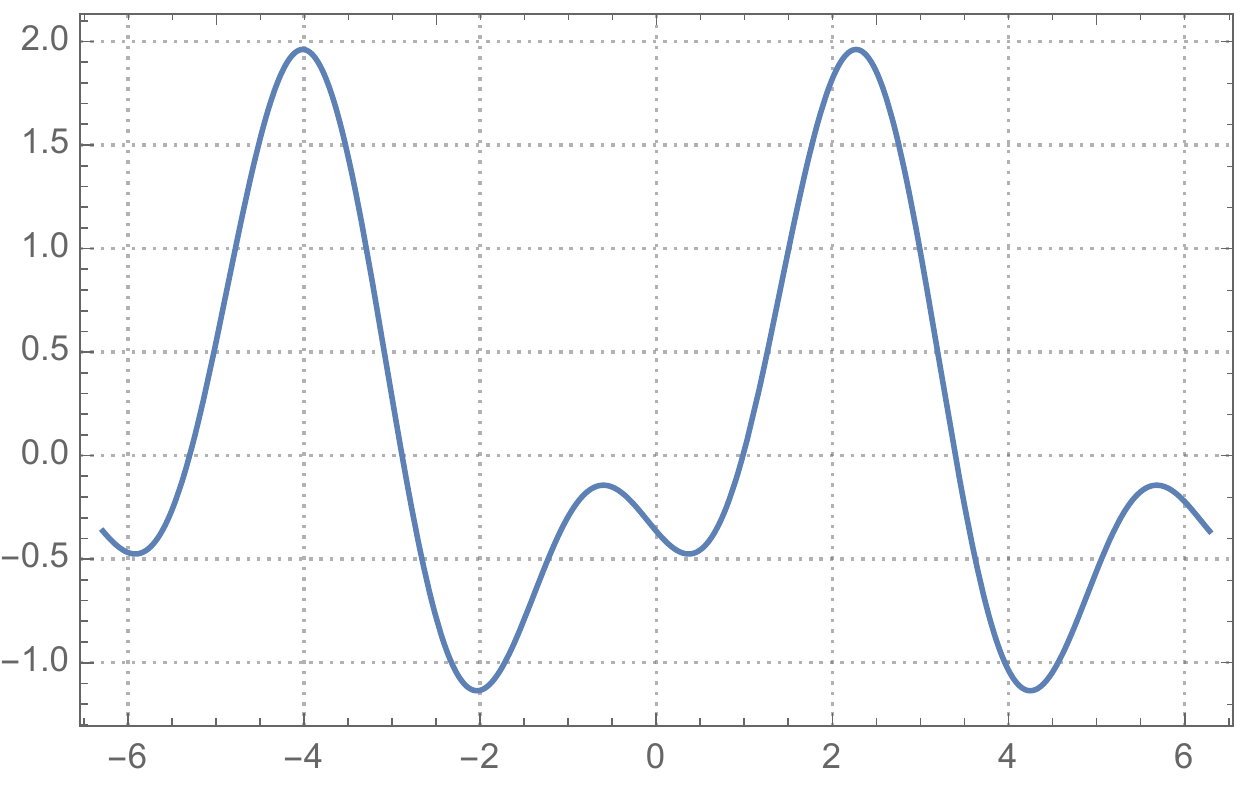}
    \caption{The first two terms.}
      \label{sg1}
  \end{subfigure}
  \begin{subfigure}
    \centering\includegraphics[width=6cm]{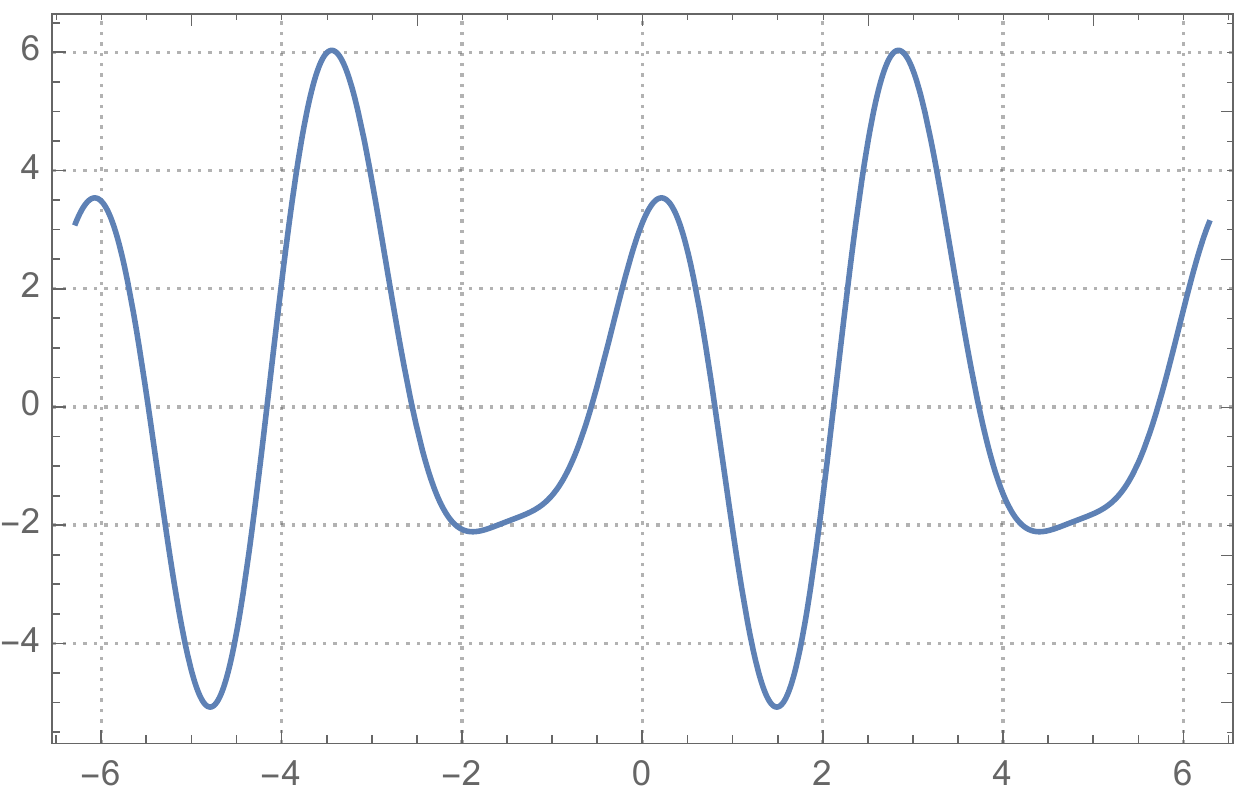}
    \caption{The first three terms.}
          \label{sg2}
  \end{subfigure}
  \begin{subfigure}
    \centering\includegraphics[width=6cm]{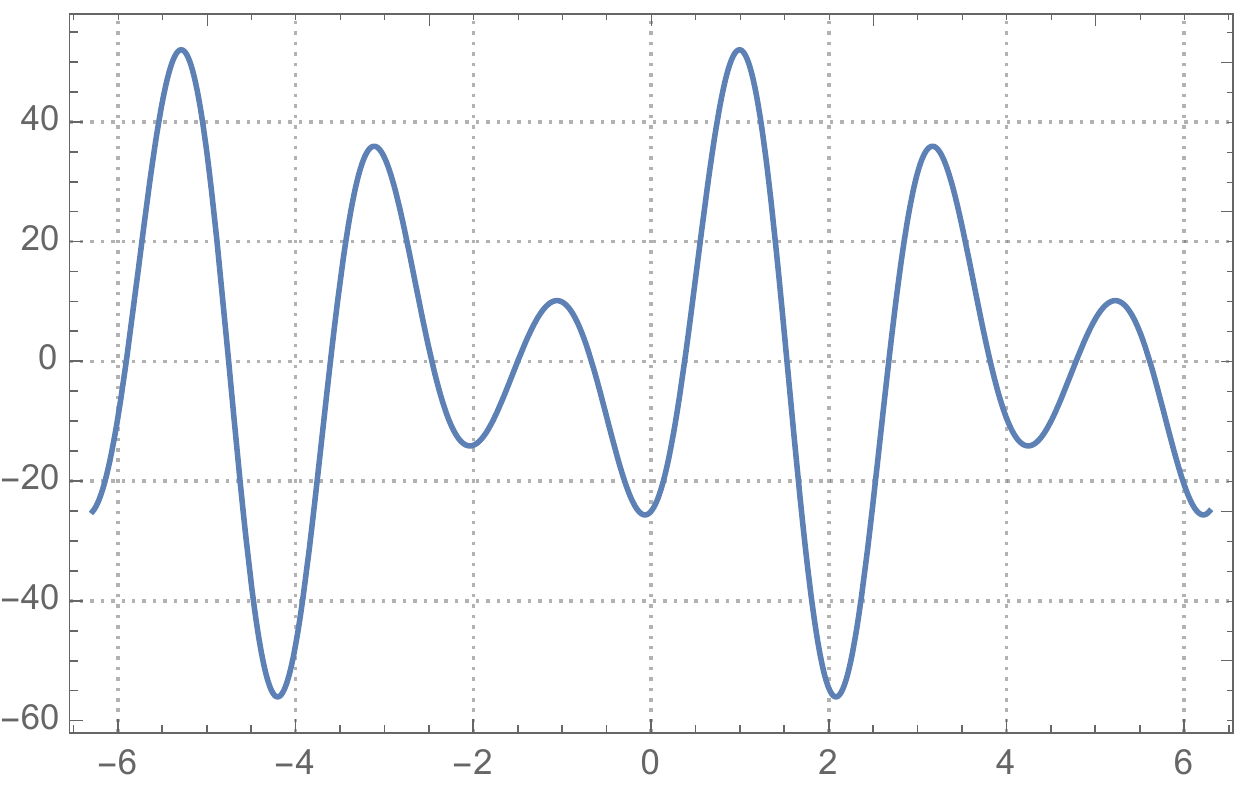}
    \caption{The first four terms.}
          \label{sg3}
  \end{subfigure}
  \begin{subfigure}
    \centering\includegraphics[width=6cm]{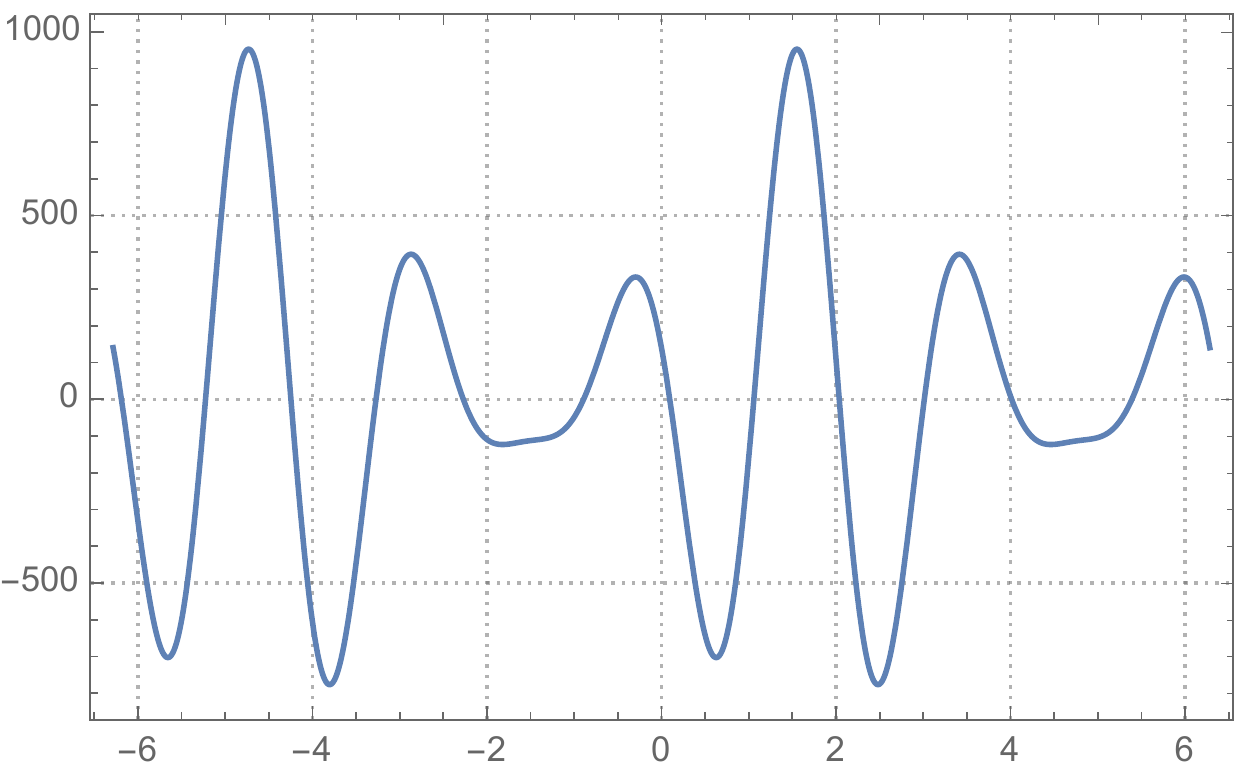}
    \caption{The first five terms.}
   \label{sg4}
  \end{subfigure}
\end{figure}

\begin{example}
This example shows the same equation 
$$u_{t} + 3\,u u_{x} + 2 \, u - \, u_{xxx} = 0,$$
where $f(x)$ of the initial solution (\ref{Gini}) will be $\sin(x)$, so that the initial guess is $v_{0}(x, t)= \sin(x)\, e^{-2t}$, then four terms of the series solution are shown in equations (\ref{term1}-\ref{term4})
\begin{align}
\label{term0}
v_{0}(x, t) &= \sin (x), \\
\label{term1}
v_{1}(x, t) &=e^{-2 t}-\frac{1}{2} (3 \sin (x)+1) \cos (x),\\
\label{term2}
v_{2}(x, t) &=e^{-2 t} \left(\frac{1}{16} e^{2 t} (-13 \sin (x)+27 \sin (3 x)+60 \cos (2 x))-3 t \cos (x)\right)+e^{-2 t},\\
\label{term3}
v_{3}(x, t)&=\frac{1}{16} \left(8 e^{-2 t} (-6 t \cos (x)+3 t (t+1) (3 \cos (2 x)-\sin (x))+2) \right) \\
                &- \frac{1}{16} \left( 27 (7 \sin (x)+19) \cos (3 x) + 417 \sin (x)+47) \cos (x)\right),\nonumber \\
\label{term4}
v_{4}(x, t)&=\frac{1}{128} e^{-2 t} \left(1728 t^3 \sin (2 x)+352 t^3 \cos (x)-864 t^3 \cos (3 x)-192 t^2 \sin (x)\right)\\
               &+\frac{1}{128} e^{-2 t}  \left(+2880 t^2 \sin (2 x)+312 t^2 \cos (x)+576 t^2 \cos (2 x)-1944 t^2 \cos (3 x) \right) \nonumber\\
               &+\frac{1}{128} e^{-2 t} \left(-192 t \sin (x)+2880 t \sin (2 x)+1763 e^{2 t} \sin (x) \right)+ \nonumber\\
               &\frac{1}{128} e^{-2 t} \left(1152 e^{-2 t} \left(-\frac{t}{2}-\frac{1}{4}\right) \sin (x)-\frac{126549}{2} e^{2 t} \sin (3 x)+\frac{6885}{2} e^{2 t} \sin (5 x)\right) \nonumber\\
               &+\frac{1}{128} e^{-2 t} \left( +72 t \cos (x)+576 t \cos (2 x)-1944 t \cos (3 x)-16332 e^{2 t} \cos (2 x) \right)\nonumber \\
               &+\frac{1}{128} e^{-2 t}\left(37908 e^{2 t} \cos (4 x) \right)+e^{-2 t},\nonumber
\end{align}
in figures (\ref{sg1}-\ref{sg4}), the plot of four approximated solutions are demonstrated, the first solution (\ref{term1}) consists of the initial term (\ref{term0}) and the first term (\ref{term1}), the second solution is the summation of the initial solution (\ref{term0}) and the two terms (\ref{term1}) and (\ref{term2}), the third solution is the summation of the initial solution and the terms (\ref{term1} ),(\ref{term2})  and (\ref{term3}), and the forth solution is the summation of the initial solution and the other four terms.
\end{example}

\section{Conclusion}
\label{conc}
The scientific result in this paper is the finding of a series solution to a new generalized KdV equation using homotopy perturbation method. In case one, the author found the solution to the equation when the linear part is assumed to be $u_{t}$, by taking different initial guesses, the solutions to the equation acted with different behaviours. In case two, the linear part has been taken to be some linear terms that belongs to the partial differential equation where its solution is easy to handle, the solution to the linear section considered as the initial guess, it contains an arbitrary function of the spacial variable in its structure. The behaviour of the solution of some numeric examples have shown a reliable convergent result as the solution approximated to a particular solution.

\bibliographystyle{unsrt}  


\newpage
\bibliography{references}
\end{document}